\documentclass[aps,prd,showkeys,nofootinbib,superscriptaddress]{revtex4-2}
\usepackage{enumerate}
\usepackage{amsmath,slashed,tensor,amssymb,epsfig,amsthm,bm,graphicx,graphics,slashed}
\usepackage[caption=false]{subfig}
\usepackage{hyperref}
\usepackage[utf8]{inputenc}
\usepackage[top=1.5cm, bottom=1.5cm, left=2cm, right=2cm]{geometry}

\usepackage{makecell}
\usepackage{microtype}
\usepackage[font=footnotesize,labelfont=bf]{caption}

\newcommand{\be}{\begin{equation}}\newcommand{\ee}{\end{equation}}
\newcommand{\bea}{\begin{eqnarray}}\newcommand{\eea}{\end{eqnarray}}
\newcommand{\brr}{\begin{array}}\newcommand{\err}{\end{array}}
\newcommand{\bit}{\begin{itemize}}\newcommand{\eit}{\end{itemize}}
\newcommand{\ben}{\begin{enumerate}}\newcommand{\een}{\end{enumerate}}

\newcommand{\bbm}{\begin{bmatrix}}\newcommand{\ebm}{\end{bmatrix}}
\newcommand{\ba}{\begin{array}}
\newcommand{\ea}{\end{array}}
\newcommand{\G}{\textbf}

\newtheorem{mydef}{Definition}
\newtheorem{Lemma}{Lemma}
\newtheorem{theorem}{Theorem}
\newcommand{\bd}{\begin{mydef}} \newcommand{\ed}{\end{mydef}}
\newcommand{\bthe}{\begin{theorem}} \newcommand{\ethe}{\end{theorem}}
\newcommand{\ble}{\begin{Lemma}} \newcommand{\ele}{\end{Lemma}}

\newcommand{\dr}{\mathrm{d}}

\def\ha{\frac{1}{2}}

\def\intx{\int \! \! \mathrm{d}^3 \textbf{x}}

\def\ph{\varphi}
\def\lan{\langle}
\def\lf{\left}

\def\non{\nonumber}\def\pa{\partial}\def\ran{\rangle}

\def\ri{\right}
\def\al{\alpha}
\def\de{\delta}\def\ep{M}

\def\la{\lambda}\def\La{\Lambda}\def\Si{\Sigma}

\def\1{{_{1}}}\def\2{{_{2}}}

\def\noHe0{:\;\!\!\;\!\!:H_e(0):\;\!\!\;\!\!:}
\def\noHm0{:\;\!\!\;\!\!:H_\mu(0):\;\!\!\;\!\!:}

\def\boldsymbol#1{{\bm #1}}

\def\lan{\langle}
\def\lf{\left}

\def\non{\nonumber}
\def\pa{\partial}\def\ran{\rangle}

\def\ri{\right}

\def\al{\alpha}
\def\de{\delta}
\def\ep{M}
\def\la{\lambda}
\def\La{\Lambda}\def\Si{\Sigma}

\def\1{{_{1}}}\def\2{{_{2}}}

\begin{document}

\title{Topologically inequivalent quantizations}
\author{G.~Acquaviva}
\email{gioacqua@utf.troja.mff.cuni.cz}

\affiliation{Institute of Theoretical Physics, Faculty  of  Mathematics  and  Physics, Charles  University, V  Hole\v{s}ovi\v{c}k\'{a}ch  2, 18000  Praha  8,  Czech  Republic.}

\author{A.~Iorio}
\email{iorio@ipnp.troja.mff.cuni.cz}

\affiliation{Institute of Particle and Nuclear Physics, Faculty  of  Mathematics  and  Physics, Charles  University, V  Hole\v{s}ovi\v{c}k\'{a}ch  2, 18000  Praha  8,  Czech  Republic.}

\author{L.~Smaldone}
\email{smaldone@ipnp.mff.cuni.cz}

\affiliation{Institute of Particle and Nuclear Physics, Faculty  of  Mathematics  and  Physics, Charles  University, V  Hole\v{s}ovi\v{c}k\'{a}ch  2, 18000  Praha  8,  Czech  Republic.}

\begin{abstract}
We discuss the representations of the algebra of quantization, the canonical commutation relations, in a scalar quantum field theory with spontaneously broken $U(1)$ internal symmetry, when a topological defect of the vortex type is formed via the condensation of Nambu--Goldstone particles. We find that the usual thermodynamic limit is not necessary in order to have the inequivalent representations needed for the existence of physically disjoint phases of the system. This is a new type of inequivalence, due to the nontrivial topological structure of the phase space, that appears at finite volume. We regard this as a first step towards a unifying view of topological and thermodynamic phases, and offer here comments on the possible application of this scenario to quantum gravity.
\end{abstract}

\vspace{-1mm}

\maketitle
\section{Introduction}
\label{sec:Introduction}

Motivated by earlier research in quantum gravity \cite{Acquaviva:2017xqi,Acquaviva:2020prd}, we explore from an algebraic point of view the apparent dichotomy between phases due to spontaneous symmetry breaking (SSB) and topological phases, and we investigate whether it is possible to reconcile these two concepts.

Topological phases of matter are a relatively novel, exotic and intensely studied field of research, see, e.g., \cite{wen2004topmatter}. They are not seen as standard thermodynamic phases, governed by Landau's theory of SSB, with an associated local order parameter, but rather as the effect of an order emerging from the nontrivial topology of fields, space or both \cite{wen2004topmatter}. On the other hand, in apparent contradiction with what just said, it has been known since long time that topologically ordered structures can be seen \textit{as the effect} of SSB, see, e.g., \cite{Umezawa:1982nv}.

Our investigation here focuses on the underlying mathematical structures that make SSB possible in the first place, hence the analysis applies equally well to field theory, condensed matter and quantum gravity. In fact, our main motivations lie in the latter, fundamental side. Indeed, the Bekenstein bound on the entropy of any physical system contained in a finite volume, could indicate that, at the most fundamental level, the Hilbert space is finite dimensional, as advocated for instance in \cite{bao2017hilbert} and \cite{carroll2019mad}. This is also our view \cite{Acquaviva:2017xqi}, which leads to a picture where both matter and space emerge from the underlying dynamics of these fermionic \cite{Acquaviva:2020prd} fundamental components, that with Feynman we call $X$ons \cite{Iorio_2019}.

This is an intriguing, and perhaps unavoidable picture, but the presence of finite degrees of freedom poses very puzzling questions: How can we have SSB in such a system? If the $X$ons indeed make everything, it is precisely their rearrangements into different phases the way we have to explain the world as it is now  \cite{Acquaviva:2017xqi}. Without SSB, how can this be? Black hole evaporation, in the first place, could not be seen as a phase transition, and the whole thermodynamics of black holes would be standing on unsound bases. Is there a way a system with finite degrees of freedom can naturally have phases? Are they, perhaps, of topological kind?

To start answering these questions, at least partially, we need to recall that at the origin of the issue here there is an old and important theorem, due to Stone \cite{Stone1930} and von Neumann \cite{vonNeumann1931} (SvN). It establishes that, for finite degrees of freedom, all representations of the canonical commutation relations (CCRs), the algebra of quantization, are equivalent up to a unitary transformation, see, e.g., \cite{Bogolyubov:1990kw}. In other words, SvN theorem establishes that a system with finite degrees of freedom only has one phase. It is then \textit{quantum infinite systems}, with an infinite number of degrees of freedom, that can have more than one phase stemming from the same Hamiltonian/Lagrangian, see, e.g., \cite{Umezawa:1982nv, Umezawa:1993yq}. That is when SSB can happen \cite{strocchi2005}.

In our case, not only the degrees of freedom are finite for $X$ons, but the associated Hilbert space is finite dimensional, and this makes the problem so hard to solve that we decided to attack it by incremental steps. The first step is to consider a simpler but related problem, that is to investigate the occurrence of phases for a system with finite degrees of freedom, but infinite-dimensional Hilbert space. Indeed SvN can also be evaded when the topology of the phase space is nontrivial, even though the degrees of freedom are finite, see, e.g., \cite{Acerbi:1992yv}. A mathematically idealized case where such ``topological escape'' from SvN appears is the quantum particle on a circle, studied extensively in \cite{Kastrup:2003fs}, see also \cite{Kastrup:2005xb} and \cite{Acerbi:1992yv}.

This appears as a promising road to pursue: to explore whether SSB can emerge in the context of finite degrees of freedom, through the \textit{topological} evasion of SvN. Nonetheless, even in this simpler setting the problem is too hard, and we need to ``simplify'' the matter even more because we first need to clarify the issues raised earlier in this Introduction, i.e., how SSB-driven phases and topological phases are related. The best setting we could envisage in order to tackle such question is that of a quantum infinite system: while we know that in such systems one can have inequivalent quantizations (that is, the inequivalent representations of the CCRs) in the infinite volume limit, we shall look also for the occurrence of \textit{topologically} inequivalent quantizations. In this way we shall have in one place both kinds of inequivalence, and this will make the comparison easier.

With this plan in mind, in what follows, we consider a quantum field theory in which \textit{vortices} are introduced through SSB implemented via the so-called {\it boson transformation} \cite{LepUme,LepManUme,Matsumoto:1973hg,Matsumoto:1975fi,Blasone:2001aj}. The presence of a vortex is formally identified by a topological defect along an axis of the configuration space.  We then show that the above is an {\it improper} transformation, meaning that it gives rise to the wanted inequivalent representations of the CCRs. This is shown by explicitly calculating the vacuum-to-vacuum amplitude and identifying the different contributions that make it vanish. When topology is trivial, to take the limit of infinite volume, by removing the infrared regulators, would be a sufficient condition for this to happen. However, the presence of the vortex introduces another, independent condition which makes the amplitude vanish at \textit{finite volume}, giving rise to the inequivalent representations of the CCRs due to a non-trivial topology of space.

Note that the problem of SSB in a finite volume was also studied in \cite{Enomoto_2006}. There, by means of the Ward--Takahashi relations, it is proved that SSB of $U(1)$ symmetry for a non-relativistic, weakly interacting Bose gas, can happen in a box of volume $V$, and that the Goldstone theorem holds. Moreover, it is shown that violations of the SvN theorem occurs even at finite volume, because ultra-violet divergences too could produce inequivalent representations of CCRs \cite{Miransky:1994vk}, due to the \emph{infinite number of degrees of freedom}.

In section II we first describe the procedure that leads to the formation of vortices in a scalar field theory.  In section III we derive an explicit expression for the vacuum-to-vacuum amplitude in the boson-transformed system in the case of a \textit{global} boson transformation, highlighting the fact that inequivalent representations of the CCRs are obtained only in the infinite volume limit. In section IV we consider a specific \textit{local} transformation that introduces a linear vortex in the system: in this case the vacuum-to-vacuum amplitude presents an additional regularization which, once removed, leads to topologically inequivalent representations of the CCRs at finite volume. In section V we discuss how such topological nontriviality is the reason for the stability of vortices at finite volume, and we propose a broader definition of SSB which encompasses both thermodynamic and topological phases. Finally, in section VI we provide our conclusions and offer a discussion on the results in the light of possible applications in the context of quantum gravitational theories.

\section{Linear vortices from boson transformations}

In order to set-up the stage and the notation, we recall here how vortices can be formed via SSB \cite{Umezawa:1993yq}. Consider a scalar field Lagrangian density of the form\footnote{Here the potential includes the mass term for $\varphi$.}
\be \label{lu1}
\mathcal{L}(x)\ = \ \pa_\mu \ph^\dag(x) \, \pa^\mu \ph(x) \, - \, V(\ph^\dag(x) \ph(x)) \, ,
\ee
invariant under the $U(1)$ rigid phase transformation
\be \label{phasetr}
\ph(x) \ \to \ e^{i \al} \, \ph(x) \, .
\ee
If this symmetry is spontaneously broken, we have (a) $\lan \ph \ran \equiv v \neq 0$, and (b) physically disjoint realizations of the system, still governed by the same action, that are the ``phases'' induced by SSB \cite{strocchi2005}. Expanding $\ph$ linearly around $v$ and using a polar decomposition $\ph(x)= (\tilde{v}+\rho(x))e^{i \, \chi(x)}$, with $\lan \rho(x)\ran=\lan \chi(x)\ran=0$, $v=\tilde{v}e^{i \chi(x)}$, the Lagrangian \eqref{lu1} takes the form \cite{ryder1996}
\be \label{lu2}
\mathcal{L}(x)\ = \ \pa_\mu \rho(x) \, \pa^\mu \rho(x) \, + \, (\rho(x)+\tilde{v})^2 \, \pa_\mu \chi(x) \, \pa^\mu \chi(x)  \, - \, V\lf(\rho(x)\ri) \, .
\ee
It can be seen from \eqref{lu2} that $\rho$ decays into $\chi$ particles, and external lines of Feynman diagrams will appear with derivatives attached.  Therefore, taking into account all-order contributions, the interacting $\ph$ can be generally expanded in terms of asymptotic fields as\footnote{Actually this expression stems from $\ph(x) \ = \ F\lf[\pa \chi^{phys} \ri] \, e^{i \, \frac{\chi^{phys}(x)}{c}}$, where $\chi^{phys}(x)$ are asymptotic $in/out$ fields or quasi-particle fields (in condensed matter physics). Here $c$ is a constant c-number, depending on renormalization constants. For simplicity, in the following, we drop the superscript $phys$ and we put $c=1$. This does not affect our considerations in any way.} \cite{NakUme, Umezawa:1982nv, Umezawa:1993yq}
\be \label{dynmap}
\ph(x) \ = \ F\lf[\pa \chi \ri] \, e^{i \,\chi(x)} \, .
\ee
Here $F$ is some functional of the asymptotic fields. Eq.\eqref{dynmap} is known as \emph{dynamical map} or \emph{Haag expansion} of $\ph$ and it is a weak mapping.  $\chi$ is the Nambu-Goldstone (NG) field, here called \emph{phason}, and it satisfies massless Klein--Gordon equation
\be
\Box \chi(x) = 0 \, .
\ee
With these, transformation \eqref{phasetr} amounts to the field translation $\chi(x) \ \to \ \chi(x) \ + \ c \, \al$, that, in order to control infrared singularities, in field theory should be actually regarded as
\be \label{btran}
\chi(x) \ \rightarrow \ \chi(x) \ + \ \al \,  f(x) \, ,
\ee
with $f$ being a square-integrable function, satisfying
\be \label{feq}
\Box f=0 \, ,
\ee
and the limit $f \to 1$ should be taken at the end of calculations. This transformation is known as \emph{boson transformation} \cite{LepUme,LepManUme,Matsumoto:1973hg,Matsumoto:1975fi,Blasone:2001aj}.
The case when $f$ is not square-integrable (e.g., when its Fourier transform does not exist), and when the limit $f \rightarrow 1$ is not taken, is particularly interesting for us, because in that case, as we shall show later, such canonical transformation leads to inequivalent representations of the CCRs, even when the infrared and ultraviolet regularizations are not removed -- in particular, when the volume is kept finite.

Such an $f(x)$ is obtained for instance when, for some point $x$,
\be \label{gmunu}
G^{\dag}_{\mu \nu}(x) \ \equiv \ \lf[\pa_\mu,\pa_\nu\ri] \, f(x) \ \neq \ 0 \, ,
\ee
$\pa_\mu f$ and $G^{\dag}_{\mu \nu}(x)$ being regular functions. It has been shown that boson transformation \eqref{btran} within the condition \eqref{gmunu} is a useful tool to describe the formation of topological defects in systems with SSB (for a general review of the subject see Refs. \cite{Umezawa:1982nv, Umezawa:1993yq}).

Some points should be remarked:
\begin{itemize}
\item
Eq. \eqref{btran} induces a local gauge transformation of $\ph$: $\ph(x) \to \ph(x) \, e^{i \, \al \, f(x)}$ . By performing  the boson transformation \eqref{btran} into the Lagrangian \eqref{lu2}, one can see that $A_\mu \equiv \pa_\mu f$ behaves as a \emph{classical} \emph{pure gauge} field.  Hence to impose Eq.\eqref{feq} is the same as to fix the \emph{Lorentz gauge} $\pa_\mu A^\mu=0$. On the other hand, the condition \eqref{gmunu} makes the gauge field nontrivial, i.e., it amounts to have a nontrivial field strength
\be \label{fmunu}
F_{\mu \nu}  = \pa_\mu A_\nu - \pa_\nu A_\mu \neq 0 \,,
\ee
at the core of the vortex, as is typical of this and other topological structures, such as instantons \cite{ryder1996} (see also the discussion of section \ref{conclusions}.  This gives the physical intuition of what is going on here, and in a moment we shall make these considerations more rigorous.
\item
Field equations are invariant under transformation \eqref{btran} if $f$ satisfies the same equation as $\chi$ (i.e. the condition \eqref{feq}). In other words
\be
\ph_f(x) \ \equiv \ F[\pa \, \chi+\pa \, f] \, e^{i \, \al \, \lf(\chi(x)+f(x)\ri)} \, ,
\ee
is an \emph{exact} solution of Heisenberg equation of motions. This is the statement of \emph{boson transformation theorem} \cite{Matsumoto:1979hs}. We remark that this result holds independently of the presence of gauge fields in the original Lagrangian (see e.g. Ref. \cite{Matsumoto:1975rp} for a proof of this theorem in the case of an Abelian Higgs-type Lagrangian), and it is strongly based on the ``gauge'' condition \eqref{feq}. Moreover, this is a non-perturbative result\footnote{A simple proof can be here sketched. Let us consider a scalar field $\Psi(x)$ satisfying the equation $\La(\pa)\Psi(x)=F[\psi(x)]$, with $\La$ being a differential operator and $F$ being a generic functional of $\Psi$. By means of Yang--Feldman equation \cite{Yang:1950vi}, this can be expanded as
\be
\Psi\lf[\Psi_0; x\ri] \ = \ \Psi_0(x) \, + \, \La^{-1}(\pa) \, F[\Psi(x)] \, ,
\ee
$\Psi_0(x)$ being an asymptotic field, satisfying $\La(\pa)\Psi_0(x)=0$. Then it is easy to see that $\La(\pa)\Psi^f(x)=F[\Psi^f(x)]$, with $\Psi^f(x) \equiv \Psi\lf[\Psi_0+f; x\ri]$, and $\La(\pa) f(x)=0$.}.
\item
It was proved that Eq.\eqref{gmunu} can be satisfied only by boson condensation of massless fields, and it then implies Eq.\eqref{feq}. In fact \cite{Matsumoto:1975rp, Umezawa:1982nv, Umezawa:1993yq}, if $(\Box+M^2)f(x)=0$, i.e. we consider boson condensation of a massive field, then
\be
\pa^\mu \, G_{\mu \nu}^\dag(x) \ = \ \lf(\Box+M^2\ri) \, \pa_\nu \, f(x) \, .
\ee
As anticipated, we assume that both $\pa_\mu f$ and $G_{\mu \nu}^\dag$ are Fourier transformable. Then
\be
\pa_\mu f(x) \ = \ \frac{1}{\Box+M^2} \, \pa^\la G_{\la \mu}^\dag(x) \, .
\ee
By deriving (contracting both members by $\pa_\mu$), we get Eq.\eqref{feq} and then $M=0$: topological defects via boson transformation can be only formed by the condensation of \emph{massless} bosons (in the present case NG bosons).
\end{itemize}

As an example, let us take $f(x) \equiv \theta$ \cite{LepUme, LepManUme,TZE197563, Matsumoto:1975rp, Umezawa:1982nv, Umezawa:1993yq}:
\be \label{bt}
\chi(x) \ \to \ \chi(x) \ + \  \al \, \theta \, ,
\ee
where $\theta$ is the azimuthal angle of a system of cylindrical coordinates, $(x_1 = r \cos \theta, x_2 = r \sin \theta, x_3 =z)$. This satisfies
\be \label{thetadelta}
\nabla^2 \theta \ = \ 0 \, , \qquad \nabla \times \nabla \theta \ = \ 2 \pi \hat{e}_3 \, \de(x_1)\de(x_2) \, .
\ee
(then $G^{\dag}_{12}(x)=-G^\dag_{21}(x)=2\pi \de(x_1) \de(x_2)$)  Note that, since $\theta$ is time independent, Laplace equation coincides with D'Alambert equation.

It can be proven \cite{Umezawa:1982nv} that the dynamical map of the Noether current associated to this $U(1)$ symmetry reads
\be
j_\mu (x)\ = \  \pa_\mu \chi(x) \, + \, O(\pa^2 \chi) \,,
\ee
hence, under (\ref{bt}), the spatial part of the Noether current transforms into
\be
\boldsymbol{j}(x) \ = \  \nabla \chi(x) \, + \, \al \, \nabla \theta \, .
\ee
By taking the vacuum expectation value (VEV), we get
\be \label{pureg}
\boldsymbol{J}(\G x) \ \equiv \ \lan \boldsymbol{j}(x)\ran \ = \ \al \, \nabla \theta \, ,
\ee
which is a relativistic analogue of the superfluid vortex ({\cite{landau1980statistical, Umezawa:1993yq}). By taking the circuitation of $\boldsymbol{J}$ around the $x_3$-axis we find
\be
Q_T \ = \ \int_{\pa \Sigma} \, \boldsymbol{J} \cdot \dr \boldsymbol{l} \ = \ \int_\Si \!\! \dr \Sigma \, \, \hat{e}_3 \cdot \nabla \times \boldsymbol{J} \ = \ 2 \pi \,  \al  \, , \label{fluxq}
\ee
where we used Stokes' theorem and Eq. (\ref{thetadelta}) (which gives $ \nabla \times \boldsymbol{J} \ = \ 2 \pi \, \al \, \de(x_1) \, \de(x_2) $).

$Q_T$ is a \emph{topological charge} \cite{Umezawa:1993yq}, and as such its conservation does not rely on the dynamics, i.e., $Q_T$ is conserved whether or not the fields are on-shell. Its general definition is given by the surface integral
\be
Q_T \ \equiv \ \ha \int \dr S^{\mu \al} \, G^\dag_{\mu \al} \,,
\ee
as can be checked for the present case. Note that there is a line (string) of singularities along the $x_3$-axis, therefore the $E(2)$ symmetry of the plane is broken. Since under the phase transformation \eqref{phasetr} also the VEV $\ph$ acquires a $\theta$ dependent phase:
\be \label{orpar}
v(\theta) \ = \ \tilde{v} \, e^{i \, \al \, \theta} \,
\ee
then, when $\theta \to \theta+ 2\pi$, one has $v(\theta) \ \to \ v(\theta) \, e^{i \, \al \, 2 \pi}$. The condition of single-valuedness for $v$ then constraints $\al \in \mathbb{Z}$, and the topological charge is quantized accordingly (vortex quantization \cite{landau1980statistical,Umezawa:1993yq}).

Let us end this section with some remarks. In the usual, semiclassical treatment of Nielsen-Olesen-Landau-Ginzburg-Abrikosov (NOLGA) vortices  \cite{Nielsen:1973cs, ryder1996, altlandsimons2010} one finds that $\varphi \to \tilde{v} e^{i \al \theta}=v(\theta)$ when $r \to \infty$, i.e., the field reaches the vacuum configuration at infinity. Therefore, the energy of a vortex diverges, e.g., in two-dimensions $E \sim \tilde{v}^2 \int_{0}^{\infty} dr 1/r \sim \tilde{v}^2 \ln r|_\infty $. The way to cure that is to make the theory local by adding a gauge field: ${\cal L} (\varphi, D_\mu \varphi, A_{\mu})$. This is a Higgs-type model, which can be viewed as a relativistic analogue of Landau--Ginzburg model for superconductors ($\varphi$ describes the condensate, $A_\mu$ the electromagnetic potential), and the vortex in this case is the persistent current. The approach we are following here is fully quantum. The classical description, as we have seen above, emerges when taking the expectation values of quantum operators. Then a natural persistent (neutral) current $\boldsymbol{J}$ dynamically emerges from the boson transformation method, by taking the VEV of the Noether current $\boldsymbol{j}$  (see Eq.\eqref{pureg}). Moreover, its value coincides with the semiclassical result for the vacuum value of vector potential ($A_\mu \to \al \, \pa_\mu \theta$, when $r \to \infty$). However, the quantum approach does not require to perform the $r \to \infty$ limit. This is particularly convenient because our result should work in the finite volume case. For the same reason here we do not care about energy divergences.

\section{Representations of the CCRs in presence of a vortex: preliminaries.}

The generator of the boson transformation \eqref{btran} in the Schr\"{o}dinger picture is
\be\label{charge}
Q \  \ = \ \intx \, f(\G x) \, \dot{\chi}(\G x)  \, .
\ee
The operator implementing such transformation is thus
\be \label{generf}
G \ \equiv \ \exp \lf(i \, \al \, Q \ri)\ = \ \exp \lf(i \, \al \, \intx \, f(\G x) \,  \dot{\chi}(\G x) \ri) \, .
\ee
so that
\be \label{bchf}
G \, \chi(\G x) \, G^\dag \ = \ \chi(\G x) \, + \, \al \, f(\G x) \, ,
\ee
Note that here $\dot{\chi}(\G x) \equiv \dot{\chi}(x)|_{t=0}$, where the reference time $t=0$ was chosen arbitrarily because the charge $Q$ is conserved.

To evaluate $Q$ explicitly we have to expand the NG field in terms of creations and annihilation operators. Since we are interested in the formation of a linear vortex along an axis, we use cylindrical coordinates.  Hence, Klein--Gordon equation reads
\be \label{ckg}
\frac{\pa^2 \chi(x)}{\pa t^2}-\frac{1}{r} \, \frac{\pa}{\pa r} \,  \lf(r \, \frac{\pa \chi(x)}{\pa r}\ri)-\frac{1}{r^2} \,\frac{\pa^2 \chi(x)}{\pa \theta^2}-\frac{\pa^2 \chi(x)}{\pa z^2} \, + \, \ep^2 \, \chi(x) \ = \ 0 \, ,
\ee
where we added a mass term for sake of generality.

Eq.\eqref{ckg} can be solved by separation of variables, as shown in Appendix \ref{kge}, and its solutions can be expanded as
\be \label{cex}
\chi(\G x)\ = \ \sum_{p,m,p_z} \, \mathcal{N} \, \lf(a_{p m p_z} \, u_{m}(p \, r, p_z \, z)+a^\dag_{p m p_z} \, u^*_{m}(p \, r, p_z \, z) \, \ri) \, ,
\ee
where we introduced the shorthand notation
\be
\sum_{p,m,p_z} \ \equiv \ \sum^{\infty}_{m=-\infty}\int^\infty_0 \dr p \, \int^{\infty}_{-\infty} \dr p_z \, ,
\ee
and we defined
\be
u_{m}(p \, r, p_z \, z) \ \equiv \ \frac{\sqrt{p}}{2\pi} \, J_m(p r) \, e^{i m \theta} \, e^{i \, p_z \, z} \,,
\ee
with $J_m(p r)$ the Bessel functions of the first kind. What we have in (\ref{cex}), then, is the scalar field expanded in cylindrical harmonics.
In order to fix the normalization we require that the equal-time CCRs,
\be \label{etcr}
\lf[\chi(\G x) \, , \,  \dot{\chi}(\G x')\ri] \ = \ i \,  \de^3(\G x-\G x') \, ,
\ee
are satisfied if
\be
\lf[a_{p m p_z} \, , \,  a^\dag_{k n k_z}\ri] \ = \ \de(p-k) \, \de_{m n} \, \de(k_z-p_z)  \, .
\ee
By using the Bessel functions closure relation\footnote{Actually, in \cite{Feshbach:2316356, Arfken:379118} this relation is proved for $\Re{n} > 1/2$. However, it holds true for every real $m$, as reported in Wolfram functions site: https://functions.wolfram.com/Bessel-TypeFunctions/BesselJ/21/02/02/0006/.} \cite{Feshbach:2316356, Arfken:379118}
\bea \label{de1}
\int^\infty_0 \! \dr p \, p \, J_n(p x) \, J_n(p y) \ = \ \frac{1}{x} \de(x-y)\, .
\eea
and the delta function representations
\bea \label{de2}
\frac{1}{2\pi} \, \sum^{+\infty}_{m=-\infty} \, e^{i m (x-y)} & = & \de(x-y) \, , \\[2mm]
\frac{1}{2\pi} \, \int^{+\infty}_{-\infty} \dr k \, e^{i k \, (x-y)} & = & \de(x-y) \, , \label{de3}
\eea
we can prove that the basis functions $u_{m}(p \, r, p_z \, z)$ satisfy the completeness relation
\be
\sum_{p \, m \, p_z} \, u_{m}(p \, r, p_z \, z) \, u^*_{m}(p \, r', p_z \, z') \ = \ \frac{1}{r} \, \de(r-r') \, \de(\theta-\theta') \,\de(z-z') \ = \ \de^3(\G x-\G x') \, ,
\ee
and the orthonormality relation
\be
\int^{+\infty}_0 \! \dr \, r \, r \int^{2\pi}_0 \! \dr \theta \int^{+\infty}_{-\infty} \! \dr z \,  u^*_{m}(p \, r, p_z \, z) \, u_{n}(k \, r, k_z \, z) \ = \  \de(p-k) \, \de_{m \, n} \, \de(k_z-p_z) \, .
\ee
By choosing $\mathcal{N}=(2 \, E)^{-1/2}$, the equal-time CCRs \eqref{etcr} follow immediately, .

Let us now come back to the charge \eqref{charge}, and use the expansion \eqref{cex} there:
\be \label{qex}
Q\ = \ -i \, \int \! r \, \dr r \, \dr \theta \, \dr z \, f(r,\theta,z) \, \sum_{p,m,p_z} \, \sqrt{\frac{E}{2(2\pi)^2}} \lf(a_{p m p_z} \, u_{m}(p \, r,p_z \, z)-a^\dag_{p m p_z} \, u^*_{m}(p \, r, p_z \, z) \, \ri) \, .
\ee
With this the generator of the boson transformation can be written in the compact form
\be \label{generator}
G\ = \ \exp\lf[\sum_{p,m,p_z} \lf(g^*_{p m p_z} \, a_{p m p_z}- g_{p m p_z} \, a^\dag_{p m p_z} \ri)\ri]\, ,
\ee
where
\be
g_{p m p_z} \ \equiv \  \, \al \, \sqrt{\frac{E}{2(2\pi)^2}} \, \int \! r \, \dr r \, \dr \theta \, \dr z \, f(r,\theta,z) \, u^*_{m}(p \, r, p_z \, z) \, .
\ee
The boson transformation \eqref{bchf}, when written in terms of the ladder operators, reads
\bea \label{bcha1}
a_{p m p_z}(g) & \equiv & G \, a_{p m p_z} \, G^\dag \ = \ a_{p m p_z} \, + \, g_{p m p_z} \, , \\[2mm]
\label{bcha2}
a^\dag_{p m p_z}(g) & \equiv & G \, a^\dag_{p m p_z} \, G^\dag \ = \ a^\dag_{p m p_z} \, + \, g^*_{p m p_z} \, .
\eea
These new ladder operators still satisfy the CCRs
\be
\lf[a_{p m p_z}(g) \, , \, a^\dag_{k n k_z}(g) \ri] \ = \ \de(p-k) \, \de_{m \, n} \, \de(p_z-k_z) \, .
\ee
Let us define the vacuum $|0\ran$ for the representation of the CCRs corresponding to $g=0$ as usual
\be
a_{p m p_z}\, |0\ran \ = \ 0 \, .
\ee
In the same way, the vacuum $|0(g)\ran$ for the representation corresponding to a nonzero $g$ is defined through
\be
a_{p m p_z}(g)\, |0(g)\ran \ = \ 0 \,,
\ee
and can be written as
\be
|0(g)\ran \ = \ G \, |0\ran \, .
\ee
One can prove that $G$ is unitarily implementable on $\mathcal{H}$ if and only if $|0(g)\ran \in \mathcal{H}$ \cite{Berezin:1966nc}. In other words, when the Hilbert spaces built on $|0\ran$ and $|0(g)\ran$ (denoted as $\mathcal{H}$ and $\mathcal{H}(g)$ respectively) are orthogonal, the canonical transformation is \emph{improper} \cite{Berezin:1966nc} and the representations of CCRs for $g=0$ and $g \neq 0$ are inequivalent.

It is then clear that the vacuum-to-vacuum amplitude
\be
\lan 0|0(g)\ran \ = \ \lan 0|G|0\ran \,,
\ee
is the quantity that indicates whether the representations of the CCRs are equivalent, hence whether SSB indeed took place.

It can be proved either by means of Baker--Campbell--Hausdorff formula \cite{Umezawa:1982nv, Umezawa:1993yq, Blasone2011} or with functional integral techniques \cite{Blasone:2017spx}, that this amplitude explicitly reads
\begin{eqnarray}
\lan 0|0(g)\ran \ = \ \exp\lf(-\ha \, \sum_{p , m , p_z} |g_{p m p_z}|^2\ri) \, .\label{g2}
\end{eqnarray}

Let us first consider the case $f(r,\theta,z)=1$. Then
it is easy to compute that
\be \label{ge}
g_{p \, m \, p_z} \ = \ 2 \pi \, \al  \, \sqrt{\frac{E}{2 \, p}} \, \de_{m \, 0} \, \de(p) \, \de(p_z) \, .
\ee
Then we notice that
\be
\sum_{p , m , p_z} g_{p \, m \, p_z}^2 \ = \ \sum_{k , n , k_z} \, \sum_{p , m , p_z} \, \de(k-p) \, \de_{m \, n} \, \de(p_z-k_z) \, g_{p \, m \, p_z} \, g_{k \, n \, k_z} \, .
\ee
We now use the Dirac-delta representations Eqs.\eqref{de1}-\eqref{de3}, obtaining\footnote{
Actually we use Eq.\eqref{de1} in the form \cite{Feshbach:2316356}
\bea
\sqrt{k \, p} \, \int^\infty_0 \! \dr r \, r \, J_n(k r) \, J_n(p r) \ = \ \de(k-p) \, .
\eea
}
\be \label{g2g}
\sum_{p , m , p_z} g_{p \, m \, p_z}^2 \ = \ \frac{1}{(2 \pi)^2} \, \int \dr z \, \int \dr \theta \int \dr r \, r \sum_{k , n , k_z} \, \sum_{p , m , p_z} \, \sqrt{k} \, \sqrt{p}  \, J_{0}(k r) \, J_{0}(p r) \, e^{i \, (m-n) \, \theta} \, e^{i \, (k_z-p_z) \, z} \, g_{p \, m \, p_z} \, g_{k \, n \, k_z} \, .
\ee
By means of the explicit expression
\be \label{sp}
\sum_{p , m , p_z} g_{p \, m \, p_z}^2 \ = \ \frac{\al^2 \, M \, V}{2} \, ,
\ee
where we used that $V = \int \dr z \, \int \dr \theta \int \dr r \, r$, the vacuum--vacuum amplitude $\lan 0|0(g)\ran$ now reads
\be \label{vtv}
\lan 0|0(g)\ran \ = \ \exp\lf(-\frac{\al^2 \, M \, V}{4}\ri) \, ,
\ee
which goes to zero when $V \rightarrow \infty$. In that limit $|0(g)\ran$ does not belong to the Fock space built on $|0\ran$ and we have unitarily inequivalent representations of CCR \cite{UmeKam}. In other words, we reproduce here the standard result that for SSB to occur we need an \textit{infinitely extended system} \cite{strocchi2005}, i.e., if the system has a \textit{finite volume} there is no SSB, no NG modes are produced to condense as a result, and the system \textit{can only describe one phase}.

Note that we introduced a mass $M$, which has to be sent to zero \textit{after} the infinite volume limit is performed. An analogous result could be achieved by using the prescription \eqref{btran} with $f(x)$ being a square integrable function \cite{Itzykson:1980rh}. This is simply a way to have a well defined mathematical procedure, that seems to convey the idea that, in usual SSB, NG bosons acquire an effective mass at finite volume\footnote{Note that here we should also deform $f$ so that $(\Box+M^2)f=0$. However, the error we commit is very small and we get the correct result \eqref{vtv} (see, e.g., Ref.\cite{UmeKam}). This discrepancy is not present for $f(r,\theta,z)=\theta$, because mass of NG field will be set to zero from the very beginning.} \cite{alfinito2002}. However, a direct calculation of the Ward--Takahashi identities, for a weakly interacting Bose gas in a box of \emph{finite} volume $V$, shows that NG modes should appear even without taking $V \to \infty$ limit, and that inequivalent representations of CCR can occur also in that case \cite{Enomoto_2006}. In this explicit example, we see a crucial aspect of why SSB-driven phases and topological phases appear to be different phenomena. The former are only possible at infinite volume, and have as distinctive feature the appearance of massless NG bosons. The latter may exist in compact, finite volumes, provided the boundary conditions are favorable, and do not require NG modes. Clearly, to search for a common ground, we need to understand the role of the infinite volume in SSB.

On this many authors have extensively commented, from different perspectives, see, e.g., \cite{Umezawa:1982nv,Umezawa:1993yq,alfinito2002,Miransky:1994vk,strocchi2005}. In the words of \cite{strocchi2005}: to have \textit{true} SSB one needs \textit{physically disjoint} realizations of the system. A physical operation on the system is necessarily a \textit{local operation} and hence, to have two physically disjoint realizations (phases) means that no local operation must be able to change one realization into another -- otherwise they are one and the same phase. Therefore, since field configurations realized at infinity cannot be changed by a local operation, a \textit{sufficient} condition for SSB is the infinite volume \cite{Miransky:1994vk}, with associated condensation of the massless NG modes.

In what follows, we shall investigate whether SSB can occur in a finite volume when a topological defect is present.

\section{Topological inequivalent representations in QFT}

In this section we set $M=0$, from the very beginning, and show that different phases occurs anyway, i.e. there is no need of the small effective mass.

In the case $f(r,\theta,z)=\theta$ one has
\be \label{gener}
G \ \equiv \ \exp \lf(i \, \al \, Q \ri)\ = \ \exp \lf(i \, \al \, \int \! r \, \dr r \, \dr \theta \, \dr z \, \theta \,  \dot{\chi}(r,\theta,z) \ri) \, ,
\ee
and
\be
g_{p m p_z} \ \equiv \  \, \al \, \sqrt{\frac{E}{2(2\pi)^2}} \, \int \! r \, \dr r \, \dr \theta \, \dr z \, \theta \, u^*_{m}(p \, r, p_z \, z) \, .
\ee
Here the integrals over $\theta$ and $z$ can be easily performed:
\bea \label{tm}
\theta_m \ \equiv \ \frac{1}{2\pi} \, \int^{2\pi}_0 \! \dr \theta \, \theta \, e^{-i m \theta}  & = &  \de_{0 \, m} \, \pi+(1-\de_{0 \, m})\frac{i}{m} \, , \\[2mm]
\frac{1}{2\pi} \, \int^{+\infty}_{-\infty} \! \dr z \, e^{-i \, p_z \, z}  & = &  \de(p_z) \, .
\eea
The radial integral
\be
f_m(p) \ = \ \int^\infty_0 \! \dr r \, r \, J_m(p r) \,,
\ee
is the \emph{Hankel} (or \emph{Fourier--Bessel}) \emph{transform} of $1$, explicitly solved in Appendix \ref{ht}
\bea
f_m(p) \ = \ \lf(\frac{m}{p^2} +\frac{\de(p)}{p}\ri) \, H(m) \, + (1-H(m)) \, (-1)^m \, \lf(\frac{\de(p)}{p}-\frac{m}{p^2}\ri) \, , \label{si}
\eea
where $H(m)$ is the Heaviside step function
\be
H(x) \ = \ \begin{cases}
       1 & \text{for x} \geq 0 \\
       0 & \text{for x} < 0 \, .
     \end{cases}
\ee
We thus find\footnote{To check that such an expression gives back Eq.\eqref{bt}, let us insert Eqs.\eqref{bcha1} and \eqref{bcha2} in the field expansion \eqref{cex}, to obtain
\begin{equation*}
\chi(\G x) \ \rightarrow \ \chi(\G x) \ + \ \pi \, \al \ + \ \,\al \, \sum_{m \neq 0} \, \int^\infty_0 \!\! \dr p \,  p \, J_m(p r) \, f_m(p) \,  \frac{ \sin\lf(m \, \theta\ri)}{m}\, .
\end{equation*}
The second term on r.h.s. comes from the $\de_{m \, 0}$ contribution in $\theta_m$. We can now perform the integral over momenta, by means of the integral \eqref{usint}, and summing over $m$, using
\begin{equation*}
\sum^\infty_{m =1}\frac{ \sin\lf(m \, \theta\ri)}{m} \ = \ \frac{\theta+2 \, k \,\pi}{2} \, - \, \frac{\pi}{2} \, , \qquad k \,\in \, \mathbb{Z} \, .
\end{equation*}
We thus get $\chi(\G x) \ \rightarrow \ \chi(\G x) \ + \ \al \, (\theta+2 \, k \, \pi)$, and noticing that the azimuthal angle $\theta$ is defined up to $2 \, k \, \pi$, we obtain Eq.\eqref{bt}.}
\be
g_{p m p_z} \ = \ \al \, \sqrt{2} \, \pi \,  p \    \theta_m \, f_m(p) \, \de(p_z)  \, .
\ee
where we used that $M=0$ and that $E \, \de(p_z)=E|_{p_z=0}$.

Now it is convenient to notice that
\bea \non
\sum_{p , m , p_z} \lf|g_{p m p_z}\ri|^2 & = & \sum^{-1}_{m,n=-\infty} \, \sum_{p , p_z , k , k_z} g_{p m p_z} \, g^*_{k n k_z} \, \de_{m \, n} \, \de(p-k) \, \de(p_z-k_z)  \\[2mm] \non
& + & \sum^{+\infty}_{m,n=1} \, \sum_{p , p_z , k , k_z} g_{p m p_z} \, g^*_{k n k_z} \, \de_{m \, n} \, \de(p-k) \, \de(p_z-k_z)
\ + \ \sum_{p , p_z , k , k_z} g_{p 0 p_z} \, g^*_{k 0 k_z} \, \de(p-k) \, \de(p_z-k_z)\\[2mm]
& = & ({\rm i}) \ + \ ({\rm ii}) \ + \ ({\rm iii}) \, .
\eea

First, the computation of the term $({\rm iii})$ ($n=0$, $m=0$) is analogue to that for $f(r,\theta,z)=1$, hence it goes to zero when $M$ is set to zero from the beginning. Then we focus on $({\rm ii})$, as the computation for $({\rm i})$ is very similar.

From Eq. \eqref{g2} we have
\bea
{\rm (ii)} && = \ \sum^{>}_{k, n, k_z, p, m, p_z} \, \frac{\al^2}{2} \, \int \dr z \, \int \dr \theta \int \dr r \, r   \, J_{0}(k r) \, J_{0}(p r) \, e^{i \, (m-n) \, \theta} \, e^{i \, (k_z-p_z) \, z} \ \frac{\de(p_z) \, \de(k_z)}{\sqrt{p \, k}}  \\
 && = \ \sum^{>}_{k, n, p, m} \, \frac{\al^2 \, h}{2} \,  \int \dr \theta \int \dr r \, r   \, J_{0}(k r) \, J_{0}(p r) \, e^{i \, (m-n) \, \theta} \ \frac{1}{\sqrt{p \, k}} \, ,
\eea
where we introduced the symbol $\sum_{}^{>}$ for sums running over positive integers, and the second expression is the result of integrating over $k_z$ and $p_z$, and on a \textit{finite} range over $z$, such that $\int^{h/2}_{-h/2} \dr z \ = \ h $. To proceed we first use the relation
\bea
\sum^\infty_{n=1} \, e^{\pm \, i \, n \, \theta} \ = \ \frac{1}{e^{\mp \, i \, \theta }-1} \, ,
\eea
which thus leads to
\bea\label{pospos}
 {\rm (ii)} \ = \ \frac{\al^2 \, h}{2} \,  \int \! \dr \theta \int \! \dr p \, \frac{1}{2 \, p^2 \, (1-\cos \theta)} \, .
\eea
The integral over $p$ can be easily regularized by introducing an infrared cut-off
\be
 \int^{+\infty}_\la \! \dr p \, \frac{1}{p^2} \ = \ \frac{1}{\la}  \, .
\ee
The regularized integral over the azimuthal angle is the most interesting:
\be \label{regthetainteg}
\ha \,\int^{2\pi-\varepsilon}_\varepsilon \! \dr \theta \, \frac{1}{1-\cos \theta} \ = \ \cot \lf(\frac{\varepsilon}{2} \ri) \, .
\ee
Here we introduced a small parameter $\varepsilon$, which geometrically corresponds to cut a thin slice around $0 = \theta = 2\pi$ direction (see Fig.\ref{fig:a1}).  Removing the cut one has $\cot(\varepsilon/2) \to +\infty$.

In the same way, by means of\footnote{We change the summation index as $n \rightarrow -n$, $m \rightarrow -m$ when they run trough negative values.}
\be
\sum^\infty_{n=1} \, e^{\pm \, i \, n \, \theta} (-1)^n \ = \ -\frac{1}{e^{\mp \, i \, \theta }+1} \, ,
\ee
one can find the contributions $({\rm i})$ $(n<0 ,m<0)$
\bea\label{negneg}
 {\rm (i)} \ = \ \frac{\al^2 \, h}{2} \,  \int \! \dr \theta \int \! \dr p \, \frac{1}{2 \, p^2 \, (1+\cos \theta)} \,,
\eea
and by a change of variables, $\theta \rightarrow \theta+\pi$, this gives the same result \eqref{regthetainteg}
\be
\ha \,\lf(\int^{2 \pi-\varepsilon}_\pi \! \dr \theta \, \frac{1}{1-\cos \theta}+\int^{3\pi}_{2\pi+\varepsilon} \! \dr \theta \, \frac{1}{1-\cos \theta}\ri) \ = \ \cot \lf(\frac{\varepsilon}{2} \ri) \, ,
\ee
regularized, once more, by means of a cut along $0 = \theta = 2\pi$.

\begin{figure}
\subfloat[]{\includegraphics[width = 3in]{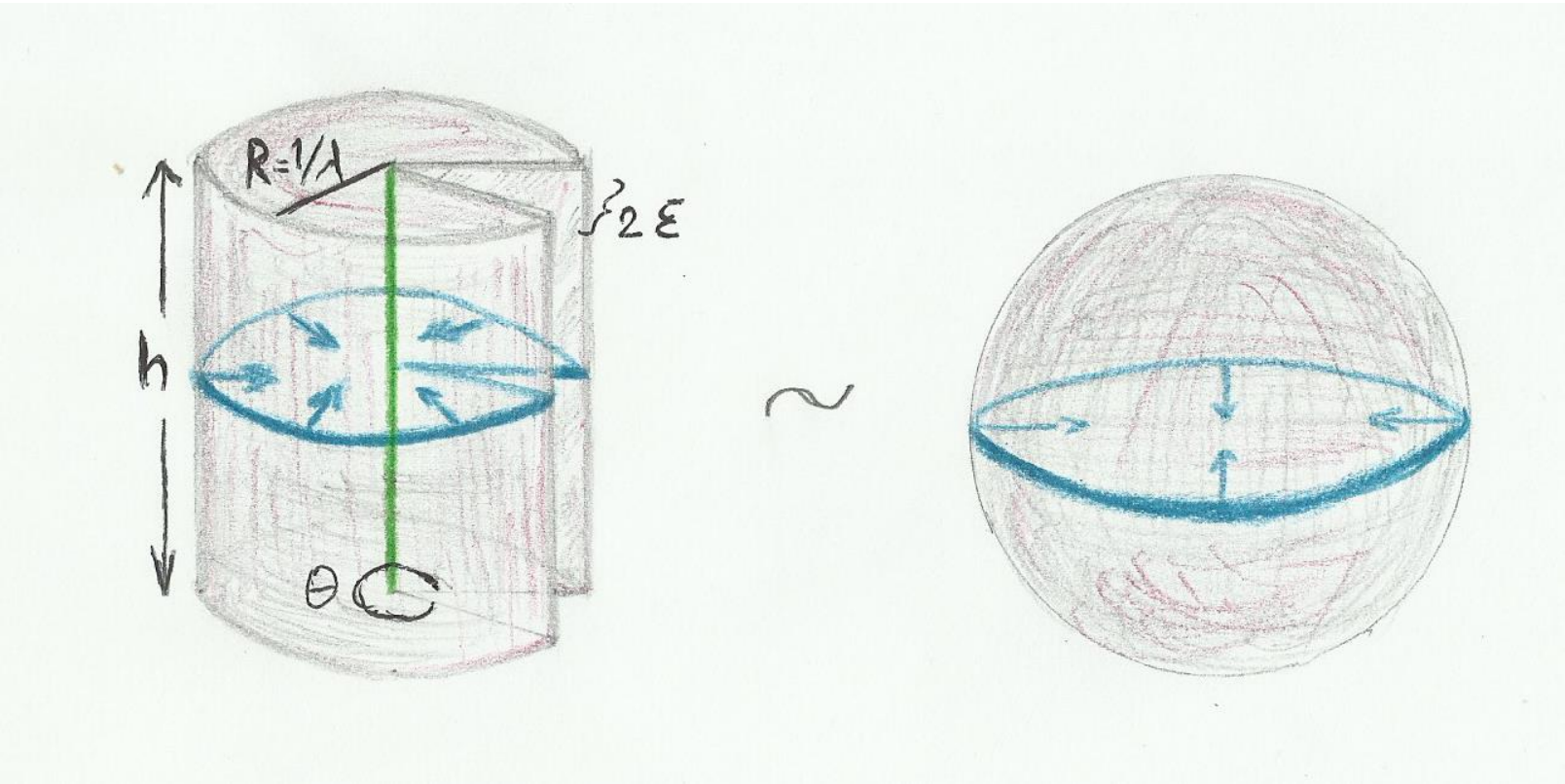}} \\
\subfloat[]{\includegraphics[width = 3.5in]{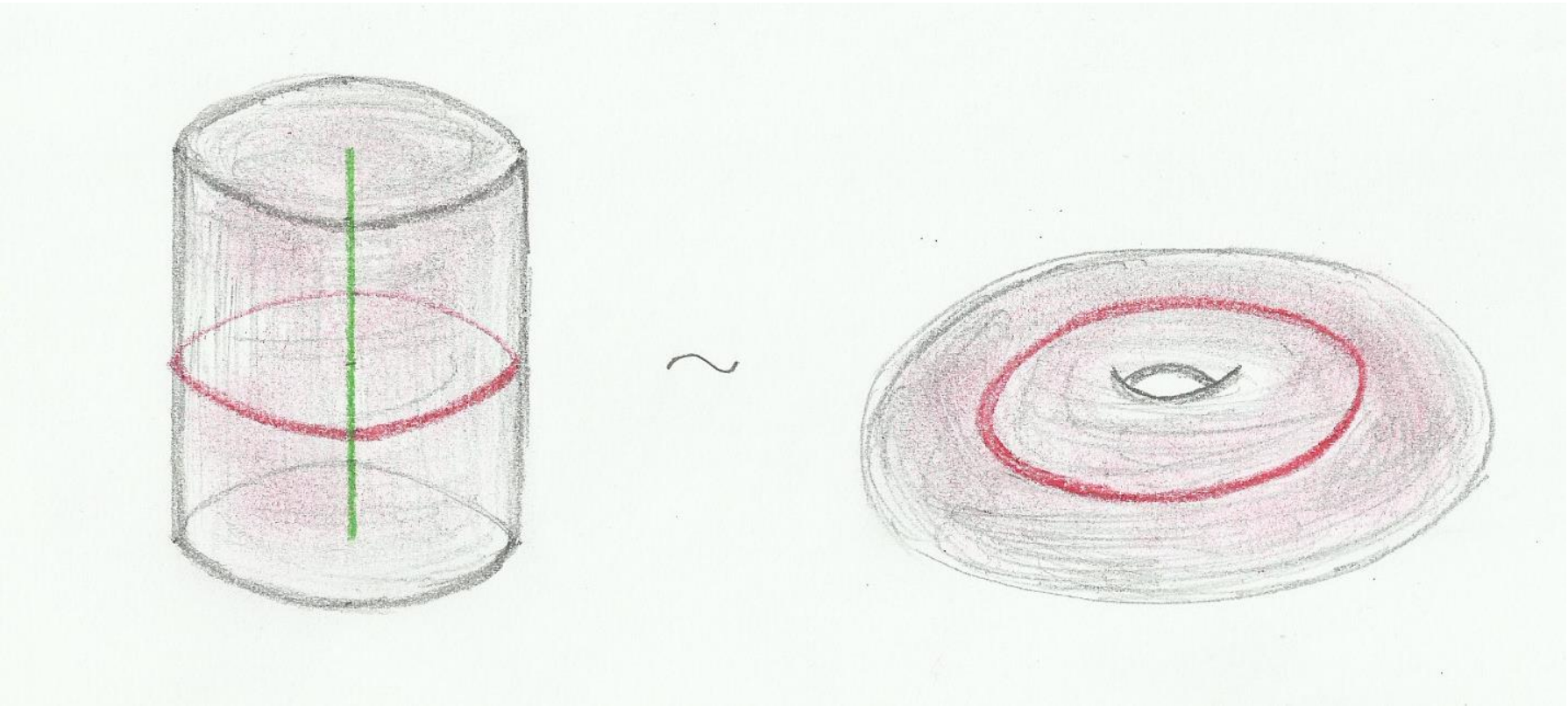}}
\caption{Figure (a) refers to the support of the fully regularized vacuum-to-vacuum amplitude, $\lan 0|0(g)\ran (\varepsilon, h, \la)$, that is the space surrounding the vortex (in green). This is a solid cylinder, of height $h$ and radius $R=1/\lambda$, with the missing slice of angle $2 \varepsilon$, homeomorphic to the ball on the right. It has trivial fundamental group, $\pi_1(B_3) = {e}$, as shown by the contractible loops (in blue). Figure (b) refers to the same space when $\varepsilon \to 0$, but $h$ and $R$ are kept finite. This is a solid cylinder with a puncture (the vortex, in green), homeomorphic to the solid torus on the right. Its fundamental group is nontrivial, $\pi_1(S^1\times D^2) = \mathbb{Z}$, as shown by the non-contractible loops (in red).}
\label{fig:a1}
\end{figure}

The final result is then
\be \label{g2top}
\lan 0|0(g)\ran ({\varepsilon, h, \la}) \ = \ \exp\lf\{-\frac{\al^2}{2} \, \frac{h}{\la} \, \cot \lf(\frac{\varepsilon}{2}\ri) \ri\}  \, .
\ee
where $h$ and $R = 1/\la$ are the height and radius of the cylinder, respectively (see Fig.\ref{fig:a1}). This is what we were looking for, and our main new result: to have in one expression both kinds of inequivalence, the standard one, driven by the thermodynamic (infinite volume) limit
\be
\lim_{h \to \infty} \lan 0|0(g)\ran ({\varepsilon, h, \la}) = 0 = \lim_{\la \to 0^+} \lan 0|0(g)\ran ({\varepsilon, h, \la}) \;,
\ee
and the inequivalence induced by the nontrivial topology
\be
\lim_{\varepsilon \to 0^+} \lan 0|0(g)\ran ({\varepsilon, h, \la}) = 0 \;.
\ee
{\it The latter inequivalence, although of topological nature and occurring also at finite volume, it is generated here through the same mechanism as the standard inequivalences emerging in the thermodynamic (infinite volume) limit.} The difference with the case $f=1$, discussed in the previous section, is in the second term of the r.h.s. of Eq.\eqref{tm}: in the case $f=1$, the sums over $m,n$ are killed by $\de_{m 0}$, $\de_{n 0}$, while in the present case we have to re-sum the infinite series, to take into account that we can wind around the singularity at $r=0$ an infinite number of times.

Let us finally notice that the introduction of an upper bound on momenta, i.e. a cut-off $\La$, so that $\la \leq p \leq \La$ does not change our previous considerations. In other words, in the present example, ultra-violet divergences do not play any role (for the inequivalence). This would suggest that topological inequivalence here appears even when the system has a \emph{finite number of degrees of freedom}.

\section{Vortices at finite volume: topology and SSB}

A topological defect of the vortex type is stable only when the field $\varphi (x)$ on the boundary (where it takes its vacuum configurations) is a map with nontrivial fundamental group. In our case
\be \label{phimap}
\varphi: \, {\rm boundary} \to U(1) \sim S^1 \,.
\ee

With reference to Fig.\ref{fig:a1},  let us then focus on the configuration space, where the vortex is a line of singularities, surrounded by a (punctured) solid cylinder. When we perform the cut along $\theta=0$, in order to regularize the integrals \eqref{pospos} and \eqref{negneg}, the resulting space is topologically equivalent to a solid sphere or ball $B_3$, as indicated in Fig.\ref{fig:a1}(a).  Therefore, the map (\ref{phimap}) is clearly trivial, $\pi_1(B_3) = {e}$, the vortex cannot stabilize and no SSB occurs.

On the other hand, see Fig.\ref{fig:a1}(b), when the $\varepsilon$-regularization is removed, the vortex is fully surrounded by the punctured solid cylinder, and the latter is homeomorphic to the solid torus, {\it i.e.} to the composition $S^1\times D^2$ of a circle and a disk. The disk is fully contractible, hence the first homotopy group of the solid torus is isomorphic to the one of the circle
\begin{equation}
 \pi_1\left( S^1\times D^2 \right) \cong \pi_1(S^1) = \mathbb{Z} \,.
\end{equation}
Therefore, keeping $h$ and $R=1/\la$ finite but performing the limit $\varepsilon \to 0$, we have $\lan 0|0(g)\ran_{\varepsilon, h, \la} \to 0$, hence the representations become inequivalent, and the vortex at \textit{finite volume} stabilizes.

In the case of $N$ vortices ($N$ line singularities in space) the bounded 3-manifold $\mathcal{M}$ is a connected sum of $N$ solid tori
\be
 \mathcal{M} = \underbrace{(S^1\times D^2)\ \#\ ...\ \#\ (S^1\times D^2)}_{N\, \text{times}}\, ,
\ee
or, equivalently, a solid $N$-torus, a ball $B^3$ with $N$ solid handles, or a so-called three-dimensional {\it handlebody of genus $g=N$}.  The fundamental group of $\mathcal{M}$ is hence the \textit{free group} of $N$ generators\cite{Nakahara}.

The question we face here is whether what we have found is \textit{true} SSB. In the vast literature on the topic SSB is often identified by an invariant dynamics (Lagrangian, Hamiltonian, equations of motion) together with a noninvariant vacuum configuration. To our knowledge, though, the true SSB is identified by the concurrence of the following five instances:
\begin{itemize}
  \item invariance of the dynamics,
  \item non-invariance of the vacuum,
  \item disjoint realizations of the dynamics (non-equivalent representations),
  \item NG (massless) bosons,
  \item thermodynamic (infinite volume) limit,
\end{itemize}
as we have discussed in the previous sections.

A way to summarize the main result of this paper is to say that in the case of the formation of vortices, $f(r,\theta,z)=\theta$, which is a prototypical topological defect, we can in fact have SSB \textit{even at finite volume}, because: 1. the disjoint realizations are obtained already in the limit $\varepsilon \to 0$, keeping $h, R$ finite, and 2. the massless NG bosons appear. That is, we have \textit{topological phases}, rather than \textit{thermodynamic phases}, but nonetheless all the other concurring instances that identify SSB are valid, as we have shown. This leads us to propose a more relaxed definition of SSB, where the last condition is replaced by a different one, that we now proceed to formulate. In this way topological and Landau's phases both will stem from SSB.

In the above recalled picture of \cite{strocchi2005} (see also, e.g., \cite{Umezawa:1982nv,Umezawa:1993yq,alfinito2002,Miransky:1994vk}), to have true SSB one needs be in the conditions that no local operation must be able to change one realization of the dynamics into another, otherwise there is always one and the same phase, even though they may appear different (i.e., associated to different vacua). Of course, since field configurations realized at infinity cannot be changed by a local operation, the infinite volume, along with the other conditions, is a sufficient condition. In our case here, we understand the inequivalence as the effect of being unable to change one topological phase into another through a local physical operation, because the operation leading from the solid torus to the solid cylinder of Fig.\ref{fig:a1}, by definition, is not a local operation. Therefore, it does not matter whether the boundary is at infinity or not, what matters is that to change the configuration of the fields at the boundary one needs a singular operation, not a local one.

Thus, the last point of the list of five concurring instances should be changed to
\begin{itemize}
  \item vacuum boundary (\textit{global}) conditions that cannot be changed by a \textit{local} operation.
\end{itemize}

Clearly, this definition includes both the thermodynamic/infinite-volume limit, regardless of the topology (corresponding, in our particular case, to $h \to \infty$ or $R \to \infty$, regardless of the fate of $\varepsilon$), and the finite volume with nontrivial topology (corresponding, in our particular case, to $\varepsilon \to 0$, and $h$ and $R$ finite). This unifies the two types of phases, thermodynamic and topological, at least in this case. Notice that, of course, it may happen that both inequivalences might be at work together, making the SSB stronger.

\section{Conclusions and outlook}\label{conclusions}

Let us conclude by commenting on the possible applications of the above described findings to the quantum gravity scenarios evoked in \cite{Acquaviva:2017xqi,Acquaviva:2020prd}, which was our main motivation in the first place. As recalled earlier, the physics of black-holes makes it natural to assume that the fundamental degrees of freedom of nature (that is those making matter, as well as space itself, $X$ons), are finite in number \cite{Acquaviva:2017xqi}. Indeed, this, together with a Pauli principle, explains the Bekenstein bound with fundamental arguments that do not assume pre-existing spatiotemporal concepts \cite{Acquaviva:2020prd}. Although fascinating, and perhaps unavoidable, this picture has many open problems, the most important being the difficulty of explaining the rich world around us made of many phases, all stemming from a single underlying dynamics, through SSB. In fact, the disjoint realizations of the $X$ons dynamics could only come from topological inequivalence, that is the other way to avoid the chains of the SvN, and topological phases are on different footings than SSB phases \cite{wen2004topmatter}.

Therefore, what we have done here is a first step (of a many-step road), towards the necessary reconciliation of these two types of phases. There are still many aspects to be understood, though. The first, is that the volume $V$, that is central to the discussion above, is the portion of $\mathbf{R}^3$ where the field $\varphi$ is immersed, while, at the fundamental level, the $X$ons make space as well as matter, hence the above picture needs a drastic rethinking. The direction to go comes from condensed matter, and is intimately related to topological defects of the dislocation and disclination kind. We are referring to the emerging (Cartan) gravity description as a theory of defects in elastic media, put forward by Kleinert \cite{kleinert1987} (see also \cite{Katanaev:1992kh}, and the exhaustive \cite{Kleinert:1989ky,kleinert3rdquant2008}), and to the recent analog model of gravity on Dirac materials \cite{Iorio:2011yz,Iorio:2013ifa,IORIO2018265,timeloopPRD2020} (see also the reviews \cite{IorioReview, Iorio:2020y5}).

In that approach, singular translations, $u^a(x)$, and singular rotations, $\omega^{a b}(x)$, in an elastic medium, play the role of our $f(x)$ in (\ref{gmunu}), giving rise to the field strengths $T^a_{\mu \nu} \sim [\partial_\mu , \partial_\nu] u^a$ and $R^{a b}_{\mu \nu} \sim [\partial_\mu , \partial_\nu] \omega^{a b}$, respectively (cf. Eq. (\ref{fmunu})). Here $T^a_{\mu \nu}$ and $R^{a b}_{\mu \nu}$ are the torsion and Riemann curvature tensors, respectively, written in the mixed notation: tangent space indices $a,b,...$ (that refer to the symmetry group to be gauged) and manifold indices $\mu, \nu, ...$.  With this in mind, to apply to that arena the results of this paper, which focuses on \textit{internal} $U(1)$ symmetry, we need to move to the gauge-gravity theory program, based on \textit{spatiotemporal} groups\footnote{As well known, this programme works particularly well in $n=3$, where Chern-Simons forms of various kinds govern the gravity theories for the different settings \cite{WITTEN198846,witten2007threedimensional,Horne:1988jf}, see also \cite{DESER1982372,GURALNIK2003222}.}, $A_\mu = e^a_\mu \mathbb{P}_a + \omega^{a b}_\mu \mathbb{J}_{a b}$: that is $ISO(n-1,1)$ for Poincar\'{e}, $SO(n,1)$ for de Sitter, $SO(n-1,2)$ for Anti de Sitter, or others, including the conformal group $SO(n,2)$, $A_\mu = e^a_\mu \mathbb{P}_a + \omega^{a b}_\mu \mathbb{J}_{a b} + f^a_\mu \mathbb{K}_a + \lambda_\mu \mathbb{D}$, see, e.g., \cite{KAKU1977304,Horne:1988jf}. Before doing so, we should consider the fate of the SvN for the \textit{discrete structures} of the quantum gravity based on $X$ons, and an interesting link between SvN and deformed CCRs with a natural discrete structure is in \cite{qSvN}.

\section*{Acknowledgements}
We would like to thank G.F. Aldi and P. Pais for useful discussions. A.I. and L.S. acknowledge support from Charles University Research Center (UNCE/SCI/013).

\appendix

\section{Klein--Gordon equation in cylindrical coordinates} \label{kge}
In this appendix we briefly show how to solve the Klein--Gordon equation \eqref{ckg} by means of separation of variables.  Firstly, we look at solutions of the form
\be
\chi(x) \ = \ \phi(\G x) \, e^{\pm i E t} \, .
\ee
This leads to
\be \label{ckgs}
-\frac{1}{r} \, \frac{\pa}{\pa r} \,  \lf(r \, \frac{\pa \phi(\G x)}{\pa r}\ri)-\frac{1}{r^2} \,\frac{\pa^2 \phi(\G x)}{\pa \theta^2}-\frac{\pa^2 \phi(\G x)}{\pa z^2} \ = \ \lf(E^2-M^2 \ri) \, \phi(\G x) \, ,
\ee
We now factorize $\phi(\G x)$ as
\be
\phi(\G x) \ = \ R(r) \, \Theta(\theta) \, Z(z) \, .
\ee
Therefore, we get:
\be \label{kgs}
-\frac{1}{R(r) \, r} \, \frac{\dr}{\dr r} \,  \lf(r \, \frac{\dr R(r)}{\dr r}\ri)-\frac{1}{\Theta(\theta) \, r^2} \,\frac{\dr^2 \Theta(\theta)}{\dr \theta^2}-\frac{1}{Z(z)} \, \frac{\dr^2 Z(z)}{\dr z^2} \ = \ E^2-M^2 \, .
\ee
We now equate the last term in the l.h.s. to a negative separation constant (we are looking at wave solutions):
\be
\frac{1}{Z(z)} \, \frac{\dr^2 Z(z)}{\dr z^2} \ = \ -p_z^2 \, ,
\ee
which is solved by $Z(z)=e^{\pm i p_z z}$. Eq.\eqref{kgs} now becomes:
\be
-\frac{r}{R(r)} \, \frac{\dr}{\dr r} \,  \lf(r \, \frac{\dr R(r)}{\dr r}\ri)-\frac{1}{\Theta(\theta)} \,\frac{\dr^2 \Theta(\theta)}{\dr \theta^2} \ = \ p^2 \, r^2 \, ,
\ee
where we defined $p^2 \equiv E^2-p_z^2-M^2$.  We further equate the second term on the l.h.s. to a negative integer (in order to enforce periodicity in $\theta$):
\be
\frac{1}{\Theta(\theta)} \,\frac{\dr^2 \Theta(\theta)}{\dr \theta^2} \ = \ -m^2 \, ,
\ee
which has solution $\Theta(\theta)\ = \ e^{\pm i m \theta}$. We are finally left with the equation:
\be \label{bde}
\lf(\frac{1}{r} \, \frac{\dr}{\dr r}+\frac{\dr^2}{\dr r^2} \ri)\, R(r) \ = \ \lf(-p^2+\frac{m^2}{r^2}\ri) \, R(r) \, ,
\ee
which is the \emph{Bessel differential equation}. The solutions are the Bessel functions (of integer order) $J_m(p r)$, $N_m(r)$ of the first and second type, respectively. We choose $J_m(p r)$, also called \emph{cylindrical harmonics}, which are regular both at $r=0$ and $r=\infty$.  In fact, we require singularities at $r=0$ to be imputed only to a singular boson transformation, as in Eq.\eqref{bt}.
\section{Hankel transform of $1$} \label{ht}
The Hankel transform of a function $a(r)$ is defined as \cite{Feshbach:2316356,DebBha}
\be \label{htrans}
A_n(p) \ = \ \int^\infty_0 \! \dr r \, r \, J_n(p r) \, a(r) \, ,
\ee
and it satisfies the relation \cite{Jackson:100964}
\be \label{jid}
a(r) \ = \ \int^\infty_0 \! \dr p \, \tilde{A}_n(p) \, J_n(p r) \, , \qquad  \tilde{A}_n(p) \ \equiv \ p \, A_n(p) \, .
\ee
We now want to evaluate the integral \eqref{htrans} in the case in which $a(r)=1$. The idea is to rewrite the required expression as:
\be
\int^\infty_0 \! \dr \, r \, J_n(p r) \,  r \ = \ \frac{\dr}{\dr \al} \lf(\int^\infty_0 \! \dr \, r \, J_n(p r) \, e^{\al r}\ri)_{\al=0}
\ee
The r.h.s can be now integrated:
\be
\int^\infty_0 \! \dr \, r \, J_n(p r) \, e^{\al r} \ = \ \frac{p^n \left(\sqrt{\alpha ^2+p^2}-\alpha \right)^{-n}}{\sqrt{\alpha ^2+p^2}} \, , \qquad n \geq 0 \, , \, p > 0 \, , \, \al \leq 0 \, .
\ee
Deriving the r.h.s. with respect to $\al$ and posing $\al=0$, one gets:
\be
\int^\infty_0 \! \dr \, r \, J_n(p r) \,  r \ = \ \frac{n}{p^2} \, , \qquad n \geq 0 \, , \, p > 0 \, .
\ee
Moreover, we know that for $n=0$  \cite{DebBha}
\be \label{intj0}
\int^\infty_0 \! \dr r \, r \, J_0(pr) \ = \ \frac{\de(p)}{p} \, .
\ee
Therefore, it is reasonable to assume that
\be \label{solvedint}
\int^\infty_0 \! \dr \, r \, J_n(p r) \,  r \ = \ \frac{n}{p^2}+\frac{\de(p)}{p} \, , \qquad n \geq 0 \, .
\ee
This result can be verified by substituting it in Eq.\eqref{jid} and using the fact that \cite{Watson}
\be
\int^\infty_0 \! \dr p \, \frac{e^{-a p}}{p} \, J_n(pr) \ = \ \frac{\lf(\sqrt{a^n+r^2}-a\ri)^n}{n \, r^n} \, ,
\ee
which, for $a=0$, reduces to
\be \label{usint}
\int^\infty_0 \! \dr p \, \frac{J_n(pr)}{p} \ = \ \frac{1}{n} \, .
\ee
Finally, we notice that, as a consequence of the well-known identity $J_n(x)=(-1)^n \, J_{-n}(x)$, we can solve the integral also for negative values of $n$. In fact
\be
\int^\infty_0 \! \dr \, r \, J_n(p r) \,  r \ = \ (-1)^n\int^\infty_0 \! \dr \, r \, J_{-n}(p r) \,  r \, .
\ee
If $n<0$, the r.h.s. is solved by Eq.\eqref{solvedint}. This concludes our proof of Eq.\eqref{si}.

As final remark we note that when $n \rightarrow 0$ the result \eqref{solvedint} reduces to the known \eqref{intj0}. This is clear for $p \neq 0$; then one must keep in mind that the limit $n \rightarrow 0$ has to be performed \emph{before} the limit $p \rightarrow 0$, giving in this way the correct result.

\bibliographystyle{apsrev4-2}
\bibliography{librarySvN}

\end{document}